# Terahertz and Infrared Spectroscopy of Gated Large-Area Graphene

*Lei Ren,[1] Qi Zhang,[1] Jun Yao,[2] Zhengzong Sun,[2] Ryosuke Kaneko,[3] Zheng Yan,[2] Sébastien L. Nanot,[1] Zhong Jin,[2] Iwao Kawayama,[3] Masayoshi Tonouchi,[3] James M. Tour,[2,4,5] and Junichiro Kono[1,6*]*

[1]Department of Electrical and Computer Engineering, Rice University, Houston, Texas 77005, U.S.A. [2]Department of Chemistry, Rice University, Houston, Texas 77005, U.S.A. Rice University, Houston, Texas 77005, U.S.A. [3]Institute of Laser Engineering, Osaka University, Yamadaoka 2-6, Suita, Osaka 565-0871, Japan, [4]Department of Computer Science, Rice University, Houston, Texas 77005, U.S.A., [5]Department of Mechanical Engineering and Materials Science, Rice University, Houston, Texas 77005, U.S.A., and [6]Department of Physics and Astronomy, Rice University, Houston, Texas 77005, U.S.A.

[*]Correspondence and requests for materials should be addressed to Junichiro Kono (kono@rice.edu).

ABSTRACT: We have fabricated a centimeter-size single-layer graphene device, with a gate electrode, which can modulate the transmission of terahertz and infrared waves. Using time-domain terahertz spectroscopy and Fourier-transform infrared spectroscopy in a wide frequency range (10-10000 $cm^{-1}$), we measured the dynamic conductivity change induced by electrical gating and thermal annealing. Both methods were able to effectively tune the Fermi energy, $E_F$, which in turn modified the Drude-like intraband absorption in the terahertz as well as the '$2E_F$ onset' for interband absorption in the mid-infrared. These results not only provide fundamental insight into the electromagnetic response of Dirac fermions in graphene but also demonstrate the key functionalities of large-area graphene devices that are desired for components in terahertz and infrared optoelectronics.

The AC dynamics of Dirac fermions in graphene have attracted much recent attention. The influence of linear dispersions, two-dimensionality, electron-electron interactions, and disorder on the dynamic conductivity, $\sigma(\omega)$, has been theoretically investigated,[1-11] whereas unique terahertz (THz) and mid-infrared (MIR) properties have been identified for novel optoelectronic applications.[12-17] For example, it has been predicted that the response of Dirac fermions to an applied AC electric field of frequency $\omega$ would automatically contain all odd harmonics of $(2n+1)\omega$, where $n$ is an integer, implying extremely high nonlinearity.[13,14] Furthermore, creation of electrons and holes through interband optical pumping is expected to lead to population inversion near the Dirac point, resulting in negative $\sigma(\omega)$, or gain, in the THz to MIR range.[12,17] While initial experimental investigations on graphene have concentrated on DC characteristics, these recent theoretical studies have instigated a flurry of new experimental activities to uncover unusual AC properties. A number of experiments have already confirmed the so-called universal optical conductivity $\sigma_0 = e^2/4\hbar$ ($e$: electronic charge and $\hbar$: reduced Planck constant) for interband transitions in a wide spectral range.[18-21] On the other hand, experimental studies of the *intraband* conductivity have been very limited,[21-24] except for successful cyclotron resonance measurements to probe Landau levels in magnetic fields.[25]

Intraband absorption is expected to increase as the Fermi energy, $E_F$, moves away from the Dirac point in either direction (*p*-type or *n*-type). On the other hand, interband absorption is possible only when the photon energy is larger than $2E_F$.[1] Thus, spectroscopic studies with a tunable carrier concentration should provide a precise determination of the location of the Fermi level, while, at the same, the capability of tuning the type and concentrations of charge carriers in graphene is desired for many electronic and optoelectronic applications. Substitution of carbon atoms in graphene by nitrogen and boron has been attempted, but this dramatically decreases the mobility by breaking its lattice structure; physically adsorbed molecules can also dope graphene, but this is not a very controllable method. Therefore, applying a controllable gate voltage to graphene to transfer carriers from a doped silicon substrate is the most commonly employed method for tuning $E_F$. By utilizing applied gate voltages, different groups have observed tunable interband optical transitions,[20,26] tunable intraband far-IR conductivity,[22] and a systematic G-band change with gate voltage in Raman spectra.[27,28]

Here, we describe our THz and MIR spectroscopy study of large-area (centimeter scale), single-layer graphene with an electrically tunable Fermi level. In a field-effect transistor configuration consisting of graphene on a $SiO_2$/*p*-Si substrate, the transmitted intensity of THz and MIR electromagnetic waves was observed to change with the gate voltage. The Drude-like intraband conductivities and the '$2E_F$ onset' of the interband transitions, monitored through time-domain THz spectroscopy (TDTS) and Fourier-transform IR (FTIR) spectroscopy, respectively, were both modulated by the gate voltage. By analyzing the spectral shape of the induced changes with appropriate models, we were able to determine the Fermi energy and scattering time as a function of gate voltage. In addition, thermal annealing under vacuum, which removes adsorbed dopant molecules, was also observed to shift the Fermi energy toward the Dirac point dramatically.

Due to the long characteristic wavelengths in THz and MIR spectroscopy, large-area samples were required for this study, as opposed to more commonly used small flakes of graphene obtained through exfoliation of graphite. Our samples were grown from a solid-state carbon source[29] – poly(methyl methacrylate) (PMMA) in the present case.[30] A thin copper foil of centimeter scale was spin coated with PMMA and then placed in the growth furnace. At a temperature between 800 and 1000°C, a reductive $H_2$/Ar gas flow was introduced under low-pressure conditions, and after ~10 minutes, a single uniform layer of large-area graphene was formed. A $CuCl_2$/HCl acid was then used to etch away the copper foil, and the remaining free-standing graphene/PMMA layer was transferred onto a lightly-doped *p*-type silicon wafer (5-10 Ω-cm) with a 300-nm thick $SiO_2$ layer. Subsequently, the PMMA layer was dissolved away by soaking the entire sample in acetone. Finally, we put gold electrodes on the four corners of the ~8 mm × 8 mm graphene film and on the back side of the Si substrate (see Fig. 1). The THz/IR wave was normal incident onto the sample, and the transmitted wave was detected and recorded as a function of gate voltage.

In the 0.1-2.2 THz (5-70 $cm^{-1}$) range, we used a typical TDTS system.[31,32] A Ti:Sapphire femtosecond laser worked as the light source to provide ~150 fs wide pulses of ~800 nm radiation at a repetition rate of 75 MHz. A (110)-oriented ZnTe crystal was used to generate coherent THz radiation via optical rectification, and a low-temperature-grown GaAs photoconductive antenna was used to detect the coherent THz radiation. In the 3-300 THz (100-10000 $cm^{-1}$) range, we used a commercial FTIR spectrometer (JASCO FT/IR-660 Plus). As Fig. 1 schematically shows, the transmission of THz/IR waves was controlled by electrically tuning the Fermi level of graphene via the gate voltage ($V_g$). A DC voltage supply was connected between one electrode on graphene and the *p*-Si substrate, and a lock-in amplifier was used at the same time to monitor the DC conductance/resistance of graphene.

Figure 2(a) shows gate-voltage-dependent, transmitted THz waveforms in the time domain, and the signal near the peak is expanded in Figs. 2(b) and 2(c). At $V_g$ = +30 V, maximum THz transmission is observed, indicating that the Fermi energy at this gate voltage is closest to the Dirac point. At all other voltages above and below +30 V, the THz transmission decreases monotonically with $V_g$, as shown in Figs. 2(b) and 2(c). Figure 2(d) shows the spectrally integrated power of the transmitted THz beam versus $V_g$ (blue circles), demonstrating that $V_g$ = +30 V (≡ $V_0$) is indeed closest to the Dirac point, and the unbiased (0 V) point (dashed line) is on the *p*-side. The DC resistance, measured *in situ*, is also plotted in Fig. 2(d) (red solid trace), showing similar gate dependence to the transmitted THz power.

After Fourier-transforming the time-domain THz data, we obtained the corresponding transmittance spectrum $T(\omega)$ in the frequency domain while treating the spectrum taken at $V_g = V_0 = +30$ V as the reference. Using standard analysis methods appropriate for thin conducting films,[33] the 2D THz conductance of the graphene sample [or, more precisely, the conductance difference between the two voltage conditions, $\Delta\sigma(\omega,V_g) = \sigma(\omega,V_g) - \sigma(\omega,V_0)$] was extracted as a function of frequency, as shown for three representative gate voltage values in Fig. 3(a). The conductance is seen to decrease with increasing frequency throughout the range of 0.3-2.1 THz (10-70 cm$^{-1}$). Here, the THz data is combined with FTIR data taken at higher frequencies (100-600 cm$^{-1}$), for which $\Delta\sigma(\omega,V_g)$ was obtained through $T(\omega,V_g) = \left[1 + \frac{\pi\alpha}{1+n_{\text{sub}}}\frac{\Delta\sigma(\omega,V_g)}{\sigma_0}\right]^{-2}$. Here $T(\omega,V_g)$ is the ratio of the transmitted signal intensity under a gate voltage to that taken at $V_g = V_0$, $\alpha$ is the fine structure constant (= 1/137), $n_{\text{sub}}$ is the real part of the refractive index of the $SiO_2$/*p*-Si substrate, obtained from a separate measurement of that substrate itself, and $\sigma_0 = e^2/4\hbar$ is the universal interband optical conductivity.

Also shown in Fig. 3(a) are theoretical fits to the data (dashed lines) to deduce the Fermi energy ($E_F$) and scattering rate ($\gamma$) using[15]

$$\sigma_{\text{intra}}(\omega,V_g) = \frac{2ie^2k_BT}{\pi\hbar^2(\omega+i\gamma)}\ln\left(e^{E_F(V_g)/k_BT} + e^{-E_F(V_g)/k_BT}\right), \qquad (1)$$

where $k_B$ is the Boltzmann constant and $T$ is the temperature (= 300 K in our experiments). As $|V_g - V_0|$ increases, $E_F$ moves away from the Dirac point, and as a result, the overall conductance increases. The extracted values for $\Delta E_F = E_F(V_g) - E_F(V_0)$ are –252 meV, –78 meV, and +31 meV, respectively, for the three spectra shown, where the negative (positive) values indicate *p*-type (*n*-type). The obtained value for $\gamma$ is ~$2 \times 10^{13}$ sec$^{-1}$ (or a scattering time of ~50 fs), which agrees with the value we obtained for the same type of graphene samples from high-field MIR cyclotron resonance measurements[34] and does not change significantly with the gate voltage.

Information on the Fermi energy can also be obtained via analysis of IR spectra, where the $2E_F$ onset sensitively changes with the gate voltage,[20,22] as shown in Fig. 3(b). Here, $\Delta\sigma_{inter}(\omega,V_g) = \sigma(\omega,V_g) - \sigma(\omega,V_0)$ is plotted for various values of $V_g$ in the 1200-9200 cm$^{-1}$ range. Each spectrum is fitted with[15]

$$\sigma_{intra}(\omega,V_g) = -\frac{ie^2}{2\pi\hbar}\int_0^\infty \frac{\sinh(x)}{\cosh[E_F(V_g)/k_BT]+\cosh(x)}\frac{\hbar(\omega+i\Gamma)/2k_BT}{x^2-[\hbar(\omega+i\Gamma)/2k_BT]^2}dx\,, \qquad (2)$$

as shown by the dashed lines, where $\Gamma$ is the broadening factor (i.e., dephasing rate) for interband transitions (assumed to be independent of photon energy and Fermi energy). The rise of $\Delta\sigma_{inter}(\omega,V_g)$ with increasing frequency on the high-frequency side can be fitted well, determining $E_F(V_g)$. On the other hand, the rise of $\Delta\sigma_{inter}(\omega,V_g)$ with decreasing frequency on the low-frequency side of each spectrum can be interpreted as the contribution of residual carriers at nominally the Dirac point voltage ($V_g = V_0 = 30$ V), likely due to the existence of electron-like and hole-like puddles. Our analysis suggests that this residual $E_F(V_0)$ value is ~175 meV. Another independent method for determining $E_F$ from the MIR spectra is to use the fact that[20,22]

$$\int \Delta\sigma_{inter}(\omega,V_g)d(\hbar\omega) = 2E_F(V_g)\sigma_0\,. \qquad (3)$$

This simple equation allowed us to deduce $E_F$ versus $V_g$ as shown in Fig. 3(c), which summarizes the gate dependence of the Fermi energy obtained through four different methods: i) fitting $\Delta\sigma_{intra}(\omega,V_g)$ in the THz by Eq. (1), ii) fitting $\Delta\sigma_{inter}(\omega,V_g)$ in the MIR by Eq. (2), iii) applying Eq. (3) to the MIR data, and iv) using a simple capacitor model[26]

$$|E_F(V_g)| = \hbar v_F\sqrt{\pi|\alpha_0(V_g - V_0)|}\,, \qquad (4)$$

where $v_F = 3 \times 10^6$ m/s is the Fermi velocity of Dirac fermions in graphene, $\alpha_0 \approx 7 \times 10^{10}$ cm$^{-2}$V$^{-1}$ for our device, and $V_0 = 30$ V. There is overall agreement between values obtained through different methods.

Finally, we also studied Fermi level tuning in a different but similar sample by thermal annealing,

which was monitored through TDTS and FTIR spectroscopy. Transmittance spectra were obtained by ratioing the transmitted signal through the graphene/substrate sample to that obtained for a reference silicon substrate with the same thickness. The real part of the 2D conductance $\sigma'(\omega)$ was then extracted, as shown in Fig. 4(a). Through the same theoretical fitting procedure as above, we determined $\gamma$ and $E_F$ to be ~$2 \times 10^{13}$ $sec^{-1}$ and –354 meV, respectively. We then annealed the sample at 200°C under vacuum for 30 minutes to remove adsorbed water and oxygen molecules. Data taken immediately after annealing (red solid line), together with its theoretical fit (red dashed line), show a significant reduction of the Drude-like intraband component, as compared with the pre-annealed case (blue solid line and blue dashed line). The result demonstrates that after annealing the Fermi level moved toward the Dirac point to a value of –170 meV from an initial value of –354 meV. This value of $E_F$ after annealing (–170 meV) is very similar to the value we obtained for $E_F(V_0)$ in the other sample, suggesting that this type of residual value of $E_F$ is typical for a large-area graphene sample, likely due to the existence of spatially inhomogeneous distributions of electrons and holes.

Environmental change can also add or reduce the physically adsorbed dopants of graphene and then shift its Fermi level. A systematic electrical gating experiment on the same graphene device was performed, and the results are shown in Fig. 4(b). When the graphene sample was exposed to air, the Dirac point was observed at ~ +80 V, as the blue line shows, which corresponds to the highest *p*-doping; after being put into a sealed box with a constant dry $N_2$ flow for 1 hour, the Diract point voltage shifted to ~ +30 V, similar to that for our TDTS experiment with the same environment. Next we placed the sample in a vacuum chamber, and under continuous pumping we observed the Fermi level to gradually move toward the Dirac point.[35,36] Finally, after vacuum pumping for 2 days, the Dirac point voltage moved to a value < +20 V, as the turquoise line shows.

In conclusion, we have studied the transmission properties of large-area graphene films in the THz and IR range. By applying an external gate voltage, we were able to electrically tune the Fermi level of graphene, which in turn modulated the transmission of THz and IR waves. In addition, thermal annealing was used to move the Fermi level toward the Dirac point. The intraband conductivity in the THz range exhibited Drude-like frequency dependence with Fermi-energy-dependent magnitude. These results provide fundamental insight into the unique AC response of Dirac fermions in graphene, while, at the same time, demonstrating promising functionalities of large-area graphene devices for critical components in terahertz and infrared applications.

ACKNOWLEDGMENTS: This work was supported by the Department of Energy (through Grant No. DE-FG02-06ER46308), the National Science Foundation (through Grant No. OISE-0530220), the Robert A. Welch Foundation (through Grant No. C-1509), and the JSPS Core-to-Core Program. J.M.T. thanks the ONR MURI (No. 00006766, N00014-09-1-1066) and the AFOSR (FA9550-09-1-0581) for financial support.

FIGURE CAPTIONS

**Figure 1:** Experimental sketch of the gated large-area graphene device fabricated, together with the incident and transmitted terahertz/infrared beams. The cartoon on the right shows the band dispersions of graphene with a gate-tuned Fermi energy.

**Figure 2:** Gate-voltage-dependent coherent terahertz wave transmission through graphene. **(a)** Transmitted terahertz waveforms through graphene under different applied gate voltages. Expanded view of the peak of the transmitted terahertz wave with varying gate voltages in the **(b)** hole regime (*p*-side) and **(c)** electron regime (*n*-side). **(d)** Transmitted terahertz power (blue open circles with lines) and DC resistance (red solid line) as a function of gate voltage.

**Figure 3:** Dynamic conductivity change of the large-area graphene sample in the terahertz/infrared frequency range induced gate voltages. Gate-induced sheet conductance change for **(a)** intraband dynamics in the terahertz regime and **(b)** interband dynamics near the $2E_F$ onset. **(c)** The absolute value of the Fermi energy $E_F$ versus gate voltage $V_g$ obtained by four different methods corresponding to Eqs. (1)-(4) (see the text for more detail).

**Figure 4:** Influence of adsorbed and environmental gas molecules on the Fermi energy in graphene detected through **(a)** terahertz/infrared spectroscopy before and after annealing as well as **(b)** DC resistance measurements under different atmospheric conditions. Inset: resistance versus gate voltage after pumping for 4 hours, 21 hours, and 48 hours.

SYNOPSIS TOC

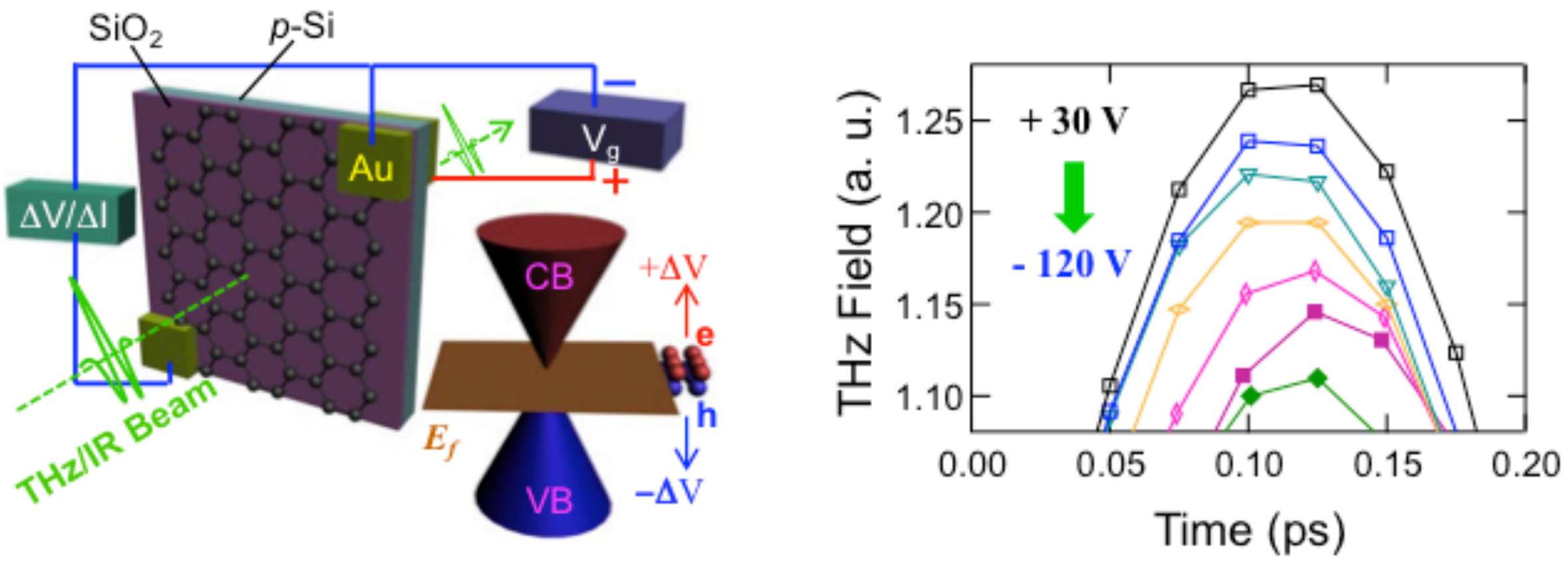

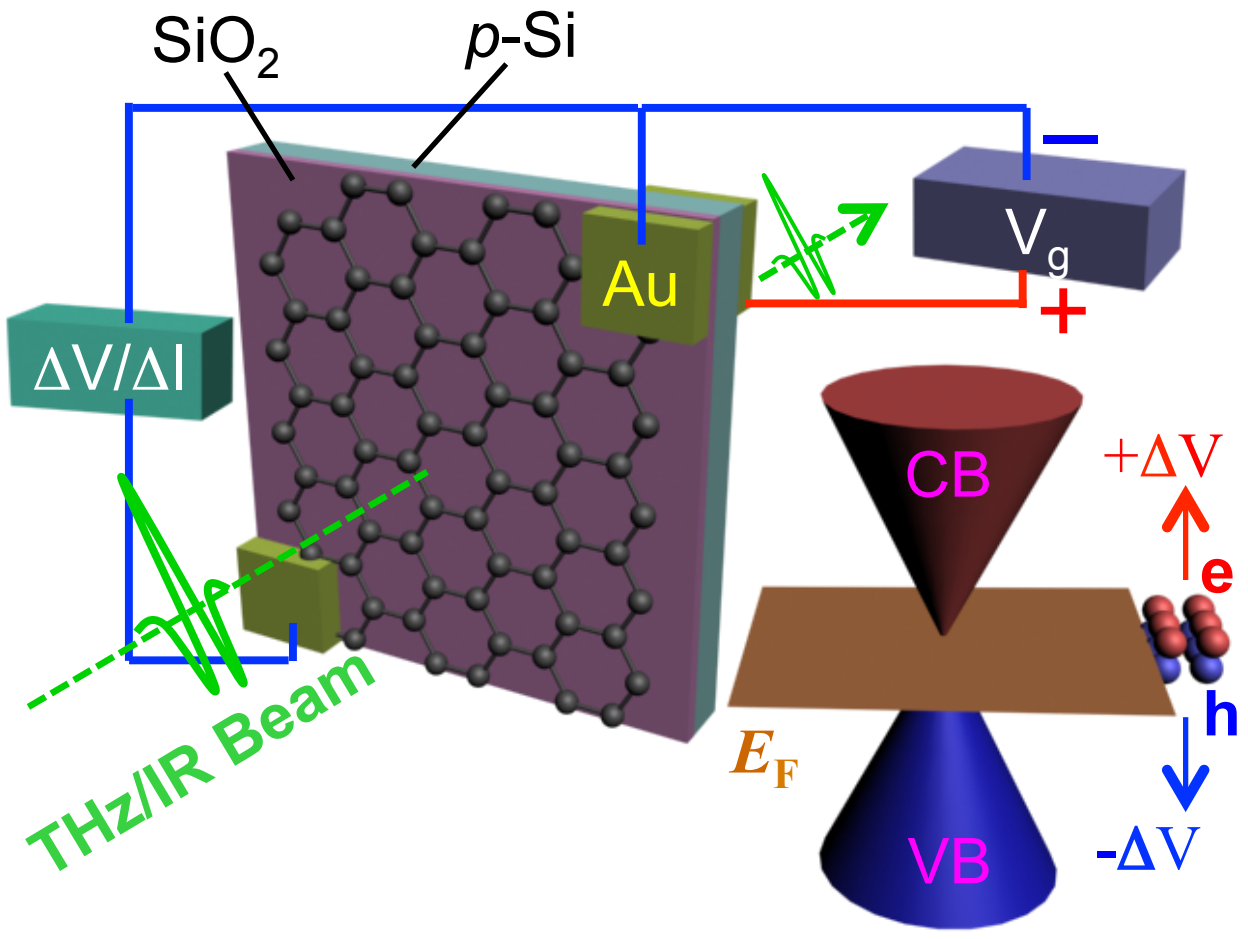

# Figure 1

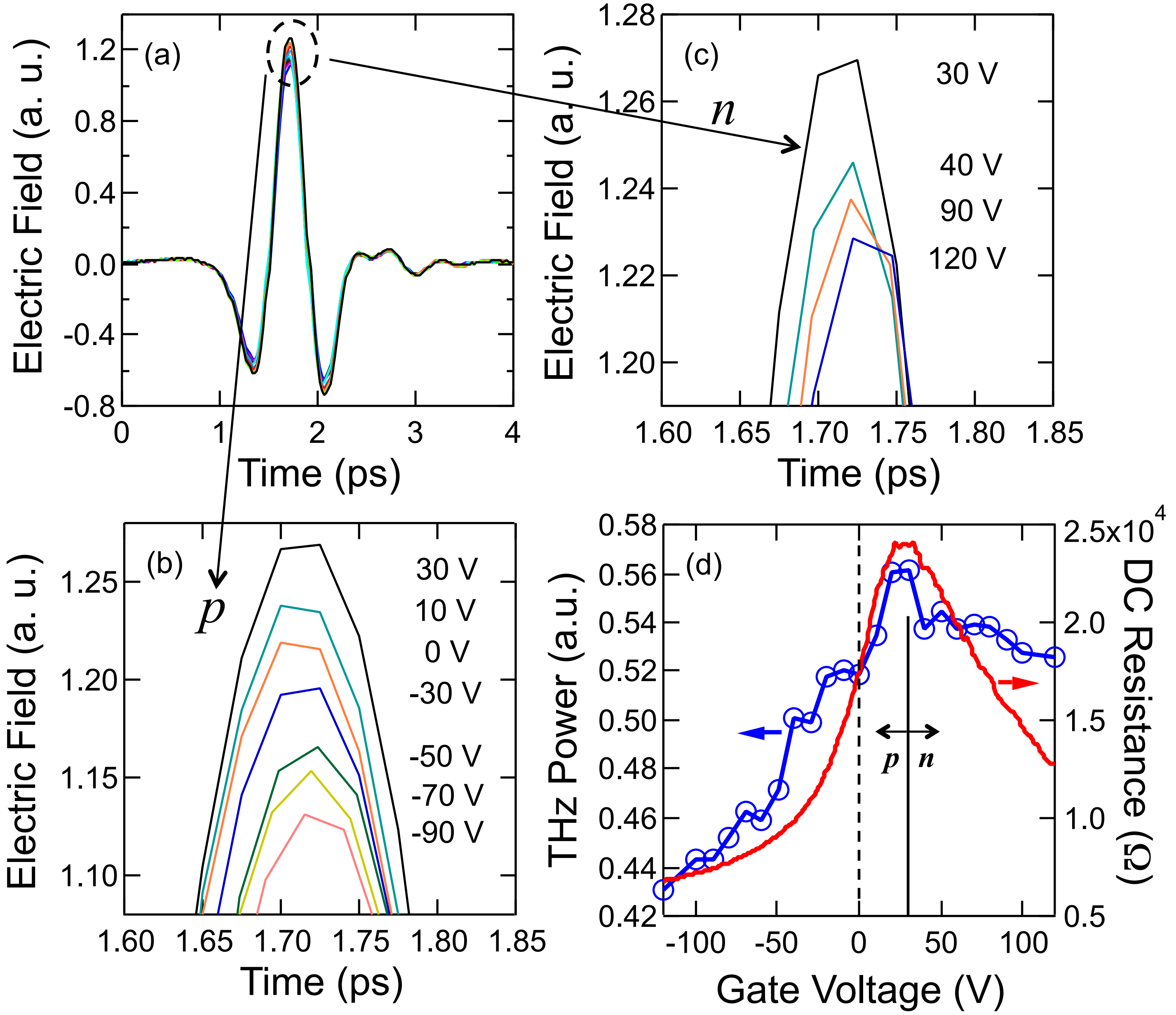

**Figure 2**

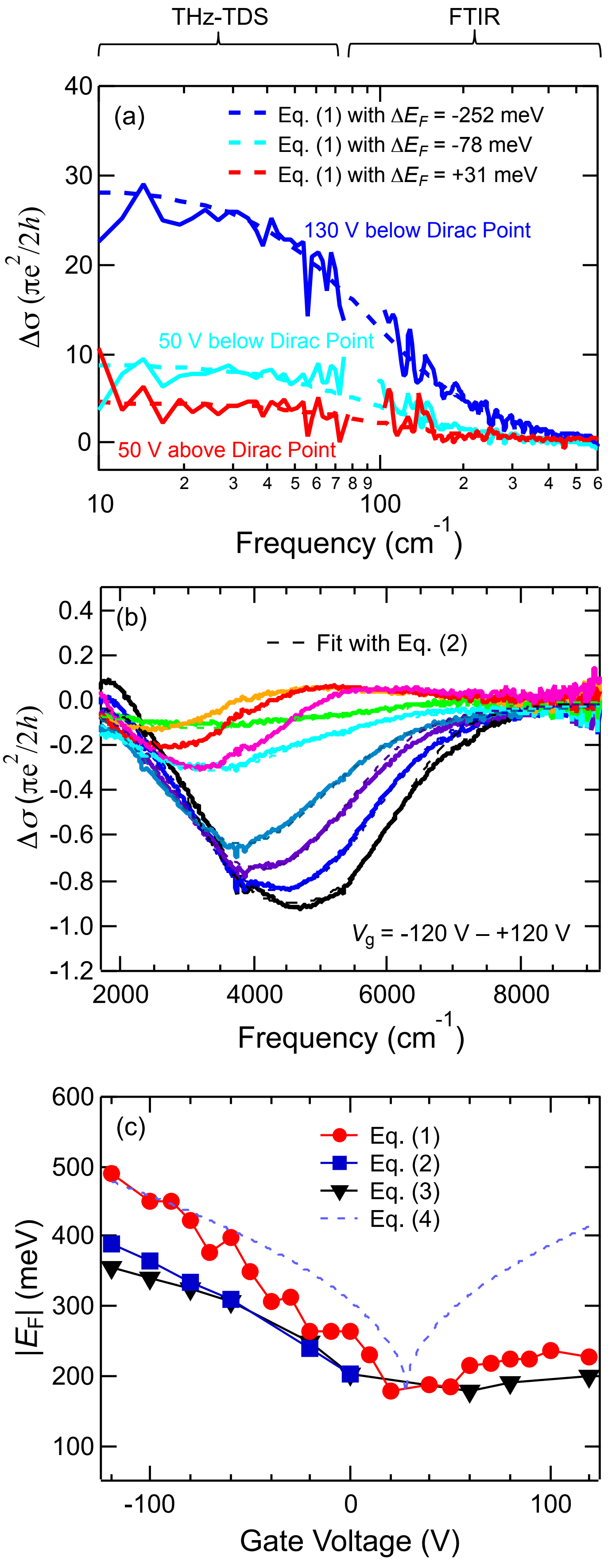
THz-TDS
FTIR
(a)
Eq. (1) with ΔE_F = -252 meV
Eq. (1) with ΔE_F = -78 meV
Eq. (1) with ΔE_F = +31 meV
130 V below Dirac Point
50 V below Dirac Point
50 V above Dirac Point
Δσ (πe²/2h)
Frequency (cm⁻¹)
(b)
Fit with Eq. (2)
V_g = -120 V – +120 V
Δσ (πe²/2h)
Frequency (cm⁻¹)
(c)
Eq. (1)
Eq. (2)
Eq. (3)
Eq. (4)
|E_F| (meV)
Gate Voltage (V)

# Figure 3

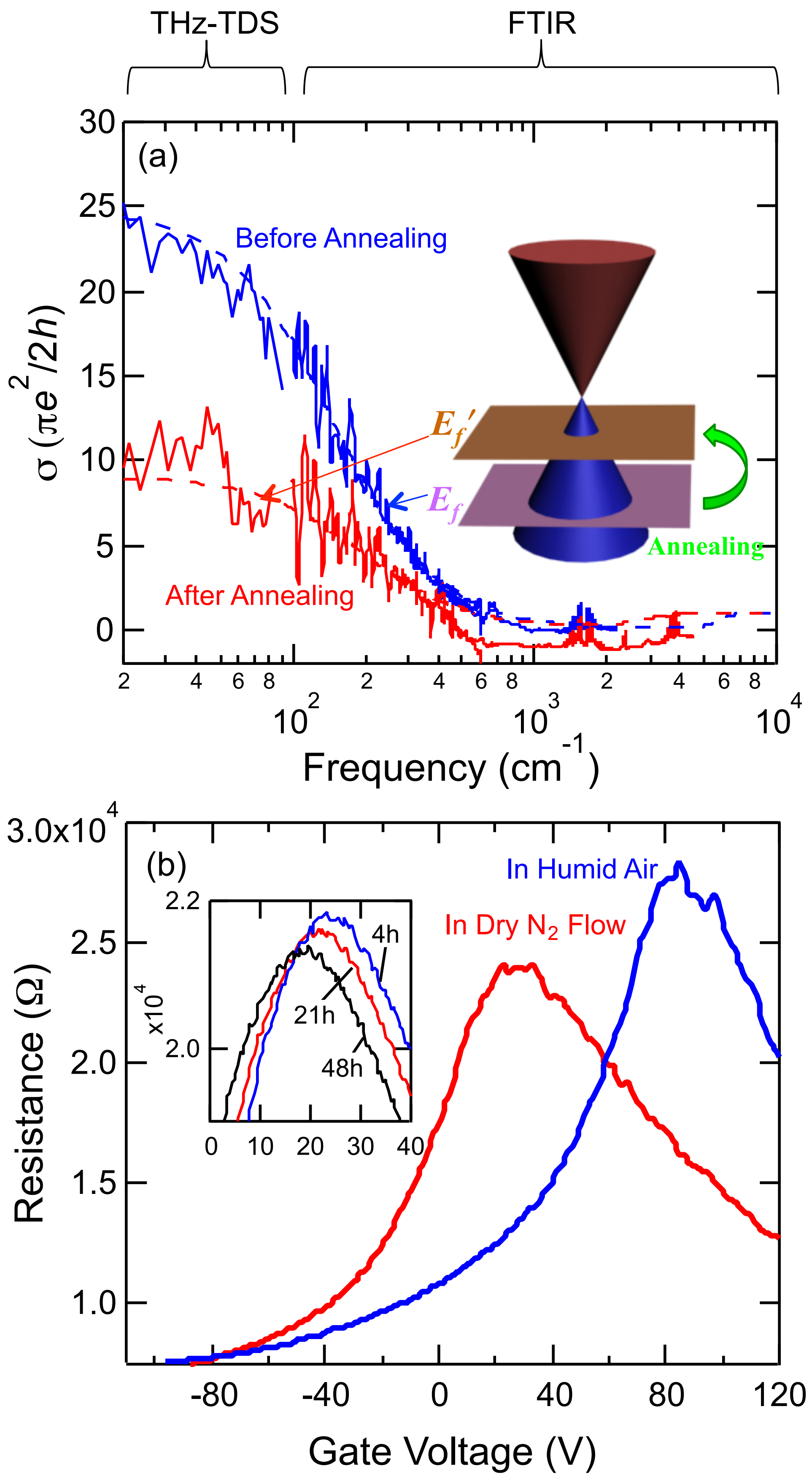

**Figure 4**